# Low Mass Companions to T Tauri Stars: a Mechanism for Rapid-rise FU Orionis Outbursts

## C.J. Clarke

*Queen Mary and Westfield College, Mile End Road, London E1 4NS.*

## D. Syer

*Canadian Institute for Theoretical Astrophysics, MacLennan Labs, 60 St. George Street, Toronto M5S 1A7, Ontario, Canada.*

**ABSTRACT**

We show that outside-in disc instabilities, which can produce rapid-rise FU Orionis outbursts in T Tauri systems, are a natural consequence of the existence of protoplanetary/protostellar companions. The protoplanetary/protostellar companion would be formed through gravitational instability on a dynamical timescale ($\sim 10^5$ y), and provided it was formed at less than a few AU, it would be swept in to $\sim 15 R_\odot$ in less than $10^5$ y. A companion of mass $10^{-2} M_\odot$ at this radius would act as a flood gate which stores material upstream and periodically releases it following the ignition of the thermal ionisation instability.

## 1 INTRODUCTION

Models of thermally unstable accretion discs have been successful in explaining the chief spectrophotometric properties of FU Orionis outbursts in young stars ( Bell *et al.* 1995, Bell & Lin 1994, Clarke, Lin & Pringle 1990). In such models, the inner disc undergoes a thermal runaway, associated with the partial ionisation of hydrogen, once its surface density achieves a critical value, $\Sigma_A$. This thermal runaway heats the inner disc (out to a few tenths of an AU) and boosts the local accretion rate by about an order of magnitude relative to its value in the outer disc, from $\sim 10^{-5} M_\odot y^{-1}$ to $\sim 10^{-4} M_\odot y^{-1}$. In this state, the modelled disc spectral energy distribution is a good fit to that observed in FU Orionis systems. The models also predict the observed line width-excitation relation (Hartmann & Kenyon 1985). Decline from outburst results from a downward thermal runaway that ensues once the surface density in the inner disc falls below another critical value.

Reproducing the observed *light curves* in FU Ori objects with such models has been more problematical however. In a disc that is isolated from its surroundings, apart from mass input at its outer edge, thermal instability first occurs at the inner edge and then propagates outwards. The resulting light curve rises slowly (on a timescale of a number of years) and in this respect resembles the slow rise system V1515 Cyg. In order to achieve a rapid rise (on a timescale of less than a year), as in V1057 Cyg and FU Orionis, it is necessary that the outburst is initiated at some larger radius (Bell *et al.* 1995, Clarke, Lin & Pringle 1990), and that the wave of runaway heating propagates inwards as well as outwards. The different temporal behaviour in this case results from the different ease with which such heating waves propagate inwards and outwards (see, for example, discussion in Lin, Papaloizou & Faulkner 1985). The triggering of the instability at larger radius can be achieved by the application of *ad hoc* density perturbations to the disc (Bell *et al.* 1995, Clarke, Lin & Pringle 1990). Whereas the resulting light curve then provides a very satisfactory fit to rapidly rising systems, the *ad hoc* nature of the perturbations is clearly undesirable, especially as no obvious mechanism has so far presented itself (see discussion in Section 4).

In this paper, we point out that the triggering of outside in outbursts (as required in rapid-rise systems) is a natural consequence of the existence of a close protoplanetary/ protostellar companion. Such a satellite can carve out an annular gap around itself, which separates material in the inner disc from that upstream of the satellite. The satellite, if sufficiently massive, causes material to accumulate upstream, whilst allowing the inner disc to drain away onto the central star. As a result, thermal runaway is first triggered just upstream of the satellite. As the disc heats up during the thermal runaway, it may be able to overwhelm the gap around the satellite and to flow into the inner disc. From this point onwards, the outburst would evolve as though the satellite were not there. After the outburst, the satellite would be able to re-create a gap in the cool disc material and the cycle would repeat itself, the satellite again controlling the location at which outburst is initiated.

In the scenario outlined above, the satellite behaves like a flood gate, which is closed during the inter-outburst period and which is suddenly opened, releasing the stored up material, when thermal instability occurs in the material upstream. In what follows, we quantify this idea, making use of the calculations of coupled disc-satellite interactions of Syer & Clarke (1995). These calculations are applicable only to rather low mass satellites ($M_s < 10^{-2} M_\odot$). Calculations in the case of more massive satellites demonstrate a qualitatively similar picture, in the sense that the satellite again generates a gap around itself that separates the inner from the outer disc (Artymowicz et al 1991, Artymowicz & Lubow 1994). The secular disc response in this case has been calculated only for the case that the disc is not subject to an external mass input, and also that the star-satellite system is not allowed to evolve in response to interaction with the disc (Pringle 1992). We stress that we concentrate on the case of lower mass satellites purely because the necessary calculations are available, but that we would expect the idea to work in a qualitatively similar manner for a binary companion more nearly equal to the primary mass.

The structure of the paper is as follows. In Section 2 we describe the coupled evolution of a satellite-disc system and calculate the radius at which thermal runaway is first triggered, comparing this radius with the requirements of time-dependent disc models for FU Orionis outbursts. In Section 3 we discuss the nature of the companion, and the constraints that are placed on its formation and evolution. Section 4 is a discussion.

## 2    DISC-SATELLITE EVOLUTION.

A satellite, mass $M_s$, orbiting a star of mass $M_\star$, clears out material in an annular gap whose extent is set by a balance between tidal and viscous effects. The gap remains open only if it is larger than both the satellite's Roche radius and the vertical pressure scale height, $H$, of the disc (Lin & Papaloizou 1985). In the case of viscous $\alpha$ discs (Shakura & Sunyaev 1973) these criteria demand that the disc aspect ratio ($H/r$, where $r$ is the distance from the central star) at the planet should satisfy:

$$H/r < \min\left[\left(\frac{q}{\alpha}\right)^{1/2}, \left(\frac{q^2}{\alpha}\right)^{1/5}\right] \qquad (1)$$

(Syer & Clarke 1995) where $q = M_s/M_\star$ and $\alpha$ is the usual viscosity parameter. Once a satellite has opened up a gap, its evolution is coupled to the disc by differential tidal torques across the gap. Depletion of the inner disc by accretion onto the star results in an unbalanced tidal torque on the satellite, which therefore spirals inwards. In the limit that the satellite is light compared with the local disc (in the sense defined below) the satellite behaves like a representative fluid element of the disc, and therefore migrates inwards on the disc's viscous timescale. A heavier satellite is unable to migrate on this timescale because of its higher inertia. As a result, material accumulates behind the satellite until it can extract enough angular momentum from the satellite to drive it inwards. The disc surface density profile adjusts itself so that its edge moves inwards with the satellite and so that angular momentum is conserved globally.

Syer & Clarke (1995) computed this coupled satellite-disc evolution in the limit of a planet relatively massive compared with the local disc, and showed that the results could be understood in terms of a simple analytical model. In this model, the ratio of the surface density just upstream of the satellite, $\Sigma_s$, relative to its unperturbed value at that point, $\Sigma_0$, depends on a single parameter $B$:

$$B = \frac{4\pi \Sigma_0 r_s^2}{M_s} \qquad (2)$$

where $r_s$ is the instantaneous radius of the satellite. $B$ is thus a measure of the satellite mass relative to the local unperturbed disc mass: the analysis presented here is applicable in the limit $B < 1$. If the relationship between $\Sigma_0$ and the accretion rate, $\dot{M}$, in a steady state disc is given by the power law:

$$\Sigma_0 \propto \dot{M}^a r^b \qquad (3)$$

then the surface density enhancement at the satellite is given by:

$$\Sigma_s = \Sigma_0 B^{-a/(1+a)} \qquad (4)$$

and the migration timescale of the satellite is related to the unperturbed local viscous timescale by:

$$t_s = t_0 B^{-a/(1+a)}. \qquad (5)$$

(Syer and Clarke 1995).

The nature of the tidal coupling between disc and satellite falls into various regimes, separated by three critical radii: $r_o$ is the radius at which the gap closure criterion (equation 1) is marginally satisfied; $r_B$, is the radius at which $B = 1$ (equation 2); and $r_A$ is the radius at which $\Sigma_s = \Sigma_A$, and thermal runaway ensues. Since $H/r$ in a steady state disc is generally an increasing function of $r$, it follows that the satellite can open up a gap when $r_s < r_o$. Inward of $r_o$, the satellite is tidally coupled to the disc and spirals inwards on the local viscous timescale. For $r_s < r_B$, the disc behind the satellite ($r > r_s$) is dammed up according to equation (4). The downstream disc ($r < r_s$) can accrete onto the central star, so that a disparity develops between the surface densities inward and outward of the satellite. If the value of $H/r$ behind the satellite at $r_A$ is boosted so as to exceed the limit given by equation (1), the gap around the satellite is overwhelmed and heating waves propagate both in and out. Provided that $r_A$ is significantly larger than the disc inner edge, the character of the ensuing outburst is 'outside-in' (and therefore rapidly rising, Bell et al 1995).

We now evaluate $r_B$ and $r_A$ for plausible models for the disc structure around young stellar objects. For the low values of $\alpha$ invoked in such discs ($10^{-4} - 10^{-3}$, Bell & Lin 1994) and for $M_s$ of a giant planet mass or above, the critical value of $H/r$ for gap closure is large, in the range several tenths to 1. Since the disc is thin between outbursts ($H/r < 0.1$, Bell et al 1995), it follows that such a satellite can open a gap at all radii. In order to calculate $r_B$, it is necessary to prescribe the radial dependence of the surface density in the unperturbed disc, which we parameterise (from the equilibrium solutions of Bell & Lin 1994) as the power law

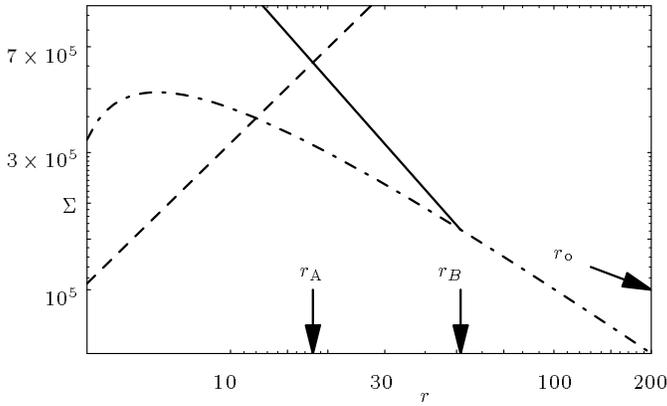

**Figure 1.** The surface density of the disc $\Sigma$ as a function of $r$ (in solar radii). The dot-dashed line is the unperturbed disc, $\Sigma_0$. The long-dashed line shows the critical surface density $\Sigma_A$ for triggering the ionisation instability. The solid line shows the effect of a satellite in banking up the disc $\Sigma_s(r_s)$. For parameters see text.

$$\Sigma_0 = 6 \times 10^4 \, \text{g cm}^{-2}$$
$$\left(\frac{r}{\text{AU}}\right)^{-0.8} \left(\frac{\alpha}{10^{-3}}\right)^{-0.8} \left(\frac{\dot{M}}{10^{-6} M_\odot \text{y}^{-1}}\right)^{0.65} \quad (6)$$

where we have used $a = 0.65$ and $b = -0.8$ (equation 3), and thus

$$r_B = 0.4 \text{AU}$$
$$\left(\frac{M_s}{10^{-2} M_\odot}\right)^{0.8} \left(\frac{\alpha}{10^{-3}}\right)^{0.7} \left(\frac{\dot{M}}{10^{-6} M_\odot \text{y}^{-1}}\right)^{-0.55} \quad (7)$$

Adopting the following radial dependence for the critical surface density for the triggering of the thermal ionisation instability (again from Bell & Lin 1994):

$$\Sigma_A = 10^7 \, \text{g cm}^{-2} \left(\frac{r}{\text{AU}}\right)^{0.9} \left(\frac{\alpha}{10^{-3}}\right)^{-0.7} \quad (8)$$

we obtain the result that the surface density at the satellite:

$$\Sigma_s = 3 \times 10^4 \, \text{g cm}^{-2} \left(\frac{r}{\text{AU}}\right)^{-1.3}$$
$$\left(\frac{\alpha}{10^{-3}}\right)^{-0.5} \left(\frac{\dot{M}}{10^{-6} M_\odot \text{y}^{-1}}\right)^{0.4} \left(\frac{M_s}{10^{-2} M_\odot}\right)^{0.4} \quad (9)$$

(equations (4) and (6)) equals $\Sigma_A$ at a radius

$$r_A = 0.06 \text{AU}$$
$$\left(\frac{M_s}{10^{-2} M_\odot}\right)^{0.2} \left(\frac{\alpha}{10^{-3}}\right)^{0.1} \left(\frac{\dot{M}}{10^{-6} M_\odot \text{y}^{-1}}\right)^{0.2} \quad (10)$$

Figure 1 illustrates the surface density profile of a steady state disc (dash-dotted line, equation 6) with $\dot{M} =$ $10^{-6} M_\odot \text{y}^{-1}$ and $\alpha = 10^{-3}$. The solid line depicts the banking up of the surface density inward of $r_B$ according to equation (4), whilst the dashed line represents the radial dependence of the critical surface density $\Sigma_A$ (equation (8)). We find that the location of $r_A$ depends rather weakly on $\dot{M}$, $\alpha$ or $M_s$ and lies in the range .02-.1AU (5-25$R_\odot$) for a satellite in the mass range $10^{-4}$-$10^{-1} M_\odot$. A satellite of Jupiter mass or above would bank up the disc behind it so that thermal instability would be triggered at a radius of around $10 R_\odot$. The value of $H/r$ once the instability is triggered is large (several tenths, Bell & Lin 1994), so that one would expect the gap to be overwhelmed once the thermal runaway is under way. The instability would then sweep into the inner disc and produce a rapid rise FU Orionis outburst.

The outburst is terminated when the disc surface density falls below a critical value, triggering downward thermal runaway to a cool state. At this point, the disc is thin again, so that the satellite can re-open a gap. The surface density in the post-outburst disc is low, due to draining of material onto the star during outburst, so that the effect on the satellite orbit is small at first. In fact, the surface density required to resume pushing the satellite downstream is given by equation (4) and is therefore, by construction, equal to $\Sigma_A$ at $r_A$. Thus the satellite, migrating inwards through tidal coupling to a disc, is 'stalled' at the radius at which thermal instability is triggered. The reason for the stalling is that at this radius material periodically overflows past the satellite and therefore relieves the torque that otherwise would drive it inwards. We therefore have the possibility that a satellite can remain at $r_A$ over a substantial period, acting as a gate for generating repeated FU Orionis outbursts. The constraints that this requirement places on the satellite are discussed in Section 3 below.

## 3 FORMATION AND EVOLUTION OF THE SATELLITE

FU Orionis outbursts are believed, on statistical grounds, to occur $\sim 100$ times over the lifetime of a T Tauri star (if all T Tauri stars undergo FU Orionis outbursts). Furthermore, the phenomenon of multiple bow shocks in Herbig-Haro objects has been tentativley interpreted as resulting from FU Orionis outbursts at intervals of $\sim 1000$ years. This leads to a period over which FU Orionis outbursts occur a few times $10^5$ years. Since the lifetime of the T Tauri phase is much longer than this, it is natural to associate FU Orionis outbursts with the early epoch of the T Tauri phase. This association is consistent with the high mass-input rates ($\sim 10^{-5} M_\odot \text{y}^{-1}$) required in disc models of the outbursts (Kawazoe & Mineshige 1993, Bell & Lin 1994) and also with the observed nebulosity, indicative of youth, around bursting objects (Kenyon et al 1987). If FU Orionis outbursts are indeed confined to young systems, then the satellites that we are postulating must have formed *quickly*.

This requirement presents no problems in the case of satellites formed through gravitational instability, either on observational or theoretical grounds. Binary systems are observed amongst the youngest T Tauri stars (age $\sim 10^5$ y, Mathieu 1994), indicating that companions can form on a timescale comparable with the free-fall time of the parent molecular cloud core. Extensive numerical simulations of collapsing rotating clouds also demonstrate the possibility of binary fragmentation on a dynamical timescale (e.g. Boss & Bodenheimer 1978, Boss 1987, Bonnell et al 1991, Bate *et al.* 1995). The low mass companions that we have considered in our calculations in Section 2 are intermediate in mass between the giant planets and stellar objects, and are too small for their formation to be resolved in the fragmentation calculations mentioned above. Such rapid formation is more problematical if the satellite is assumed to form by prior agglomeration of a rocky core, where timescales of $\sim 10^8$ years are usually considered necessary (Bodenheimer & Pollack 1986, Nakagawa et al. 1983). This latter estimate is however inconsistent with the low gas masses observed around pre-main sequence stars (Zuckerman et al 1995) which implies that rapid formation (as in the models of Lissauer 1987 and Barge & Sommeria 1995) is required. We stress that although the satellites we postulate overlap the giant planets in mass, they do not necessarily resemble them in structure and formation history: the satellites considered here are ultimately accreted by the central star and so do not survive the pre-main sequence stage.

The second requirement of our model is that the companions, once formed, can find their way to a radius $r_A$ ($\sim 0.1$AU) within $\sim 10^5$ years. This requirement is satisfied for objects that form at sufficiently small radii that disc-satellite coupling can bring them in on the required timescale. For $r_s > r_B$ (equation 7) the satellite spirals in on the disc's viscous timescale

$$t_0 = 4 \times 10^4 \text{y}$$
$$\left( \frac{r}{\text{AU}} \right)^{1.2} \left( \frac{\alpha}{10^{-3}} \right)^{0.8} \left( \frac{\dot{M}}{10^{-6} M_\odot \text{y}^{-1}} \right)^{-0.35} \quad (11)$$

as derived from the equilibrium solution (6). For $r_s < r_B$ the migration timescale is given by equation (5): although this implies that $t_s$ increases relative to $t_0$ at small radii, the steep inward decrease of $t_0$ means that the largest timescales are still encountered at large radii. Thus, from equation (11), the requirement that the satellite can be swept into a radius $r_A$ within $\sim 10^5$ years translates roughly into a requirement that it be formed at radii of a few AU or less.

The final requirement of our model is that the satellite is stable against tidal disruption at $r_s = r_A$, which effectively means that it must already have undergone the hydrodynamic collapse associated with dissociation of molecular hydrogen at its core. This point is reached on a Kelvin-Helmholtz timescale when the central temperature reaches 2000 K. (Note that a satellite that is disrupted can only give rise to a single FU Orionis outburst, caused by its mass accreting onto the central object. Whereas one might postulate a wealth of protoplanetary/protostellar companions so as to explain FU Orionis outbursts as multiple satellite disruptions, we concentrate here on the more economical solution of using a single satellite).

From models of the formation of brown dwarfs (Burrows *et al.* 1993), we find that an object of mass $\sim 0.01 M_\odot$ at an age of around $10^5$ years is more dense, on average, than a T Tauri star. Such an object would fill its Roche lobe only if brought close to the surface of the T Tauri star, and thus is not susceptible to tidal disruption at $r_s = r_A$. A similar conclusion can be reached from models of the formation of Jupiter through gravitational collapse (Bodenheimer et al 1980). Note that before hydrodynamic collapse the protoplanetary/protostellar companion must be stable against tidal disruption at its site of formation. This condition is satisfied in models of formation of giant planets through gravitational instability at about 5AU(e.g. Bodenheimer *et al.* 1980).

Following hydrodynamic collapse, which occurs on a timescale of around a year, the resulting structure is compact, with a high central temperature ($\sim 30,000$ K). It is notable that the high central temperature of protoplanetary/protostellar companions would then from complete evaporation during the FU Orionis outburst, when immersed for periods of $\sim 100$ years in gas at several times $10^4$K. Partial evaporation of the outer layers would be expected however: the temperature inversion induced by heating the satellite from the outside would tend to supress convective cooling and thus puff up the outer layers, as energy released from the satellite's gravitational contraction would be unable to escape (Cameron *et al.* 1982). Substantial (i.e. order unity) changes in the satellite's radius would however requires that such immersion lasted a Kelvin-Helmholtz timescale. Thus the brief interludes of immersion envisaged here would be expected to have only a modest influence on the tidal stability criterion of the satellite.

## 4  CONCLUSIONS

We argue that, in the scenario outlined above, the presence of the satellite sets up the disc in a state suitable for the generation of outside-in outbursts, as required in rapid rise FU Orionis systems. This argument is based on the detailed investigation of Bell *et al.* (1995) who found that essentially any perturbation that destabilised the disc at radii significantly greater than its inner edge gave rise to outside-in outbursts of the required form. Past conjectures about the origin of such perturbations have involved the effects of a massive binary companion on an eccentric orbit that periodically disturbs the disc (e.g. Kenyon et al 1988, Clarke et al 1990, Bonnell & Bastien 1992, Bell *et al.* 1995), but the absence of observed companions at suitable radii is a problem for such models. We note that in the present model the protoplanetary/protostellar companions, required as 'gates' would not be detectable spectroscopically in currently outbursting systems, since the central star is swamped by disc emission during FU Ori outbursts. The detection of suitable protoplanetary/protostellar companions (which induce radial velocity shifts of the central star of less than $1 km/s$) in T Tauri stars may however be within the reach of a new generation of high precision spectrometers.

It is not clear at present what fraction of FU Orionis outbursts are of the rapid rise variety. Two out of the six confirmed FU Orionis systems are in this category, but this statistic needs to be interpreted with caution given the observational bias towards identifying systems whose brightness has changed dramatically (Bell *et al.* 1995). Although this diversity of light curve behaviour may alternatively result from intrinsic differences in disc properties, it is clearly undesirable to have to postulate variations in such fundamental properties as the viscosity parameter, for example, from object to object.

In the present picture, as in the scenario proposed by Bell & Lin (1994), the epoch of FU Orionis outbursts is terminated when the mass input to the outer disc falls below the critical value at which the thermal ionisation instability can be triggered somewhere in the disc. As the mass supply dwindles, $r_A$ also falls (equation 10) so that the satellite is pushed to progressively smaller radii: depending on its detailed equation of state and structure it may or may not be tidally disrupted before finally being accreted by the central object. (The issue of dynamical tidal instability however requires further exploration (analogous, for example to that of Hjellming & Webbink 1987 in the case of composite stars) for objects with the complex equations of state of protoplanetary/protostellar companions). If, on the other hand, the disc surface density is abruptly reduced (due, for example, to the clearing action of stellar winds) then such a satellite may avoid being accreted, being instead stranded at its instantaneous location when the disc was dispersed. Detailed calculations of tidally coupled thermally unstable discs are required in order to ascertain whether satellites are likely to be stranded in this way, and thus whether such objects - similar to that recently discovered in 51 Pegasus (Mayor and Queloz 1995) - should be commonly found in Weak Line T Tauri stars and post- T Tauri stars.